Machine-Learning-Guided Prediction Models of Critical Temperature of Cuprates

Donggun Lee[1], Daegun You[2], Dongwoo Lee[2], Xin Li[3], and Sooran Kim[1*]

[1]Department of Physics Education, Kyungpook National University, Daegu, 41566, Korea

[2]School of Mechanical Engineering, Sungkyunkwan University, Gyeonggi-do, South Korea

[3]John A. Paulson School of Engineering and Applied Sciences, Harvard University, Cambridge, Massachusetts 02138, USA

*Corresponding author: sooran@knu.ac.kr

**Abstract**

Cuprates, a member of high-$T_c$ superconductors, have been on the long-debate on their superconducting mechanism, so that predicting the critical temperature of cuprates still remains elusive. Herein, using machine learning and first principle calculations, we predict the maximum superconducting transition temperature ($T_{c,max}$) of hole-doped cuprates and suggest the explicit functional form for $T_{c,max}$ with the root-mean-square-error of 3.705 K and the coefficient of determination $R^2$ of 0.969. We employed two machine learning models; one is a parametric brute force searching method and another is a non-parametric random forest regression model. We have found that material dependent parameters such as the Bader charge of apical oxygen, the bond strength between apical atoms, and the number of superconducting layers are important features to estimate $T_{c,max}$. Furthermore, we predict the $T_{c,max}$ of hypothetical cuprates generated by replacing apical cations with other elements. When Ga is an apical cation, the predicted $T_{c,max}$ is the highest among the hypothetical structures with 71, 117, and 131 K for one, two, and three $CuO_2$ layers, respectively. These findings suggest that machine learning could guide the design of new high-$T_c$ superconductors in the future.

Understanding the material dependence of superconducting temperature ($T_c$) has been a long-standing subject of importance in the condensed matter physics community. However, this becomes especially challenging for the high $T_c$ cuprates, as their underlying mechanism of superconductivity, despite intensive experimental and theoretical studies, still remains elusive over 30 years since the discovery of $La_2CuO_4$ [1]. All cuprate superconductors share the common characteristics of the two-dimensional $CuO_2$ superconducting layer and strong electronic correlations [2], while their experimental maximum superconducting transition temperatures ($T_{c,max}$) vary by an order of magnitude from the La family to the Hg family. This suggests that out-of-$CuO_2$-plane effects should also be integrated into the superconducting mechanism. Furthermore, complex physical phenomena such as abnormal phonons, charge density wave, and pseudo gap state bring continuous interest and discussion to copper-based superconductors [2–8].

Several material dependent parameters that have strong correlations with $T_c$ of cuprates have been suggested. Experimentally, the $T_c$ increases with the number of $CuO_2$ layers up to three and decreases beyond four [9–11]. The apical oxygen height ($d_A$), the distance between the apical oxygen and in-plane copper, has been considered to have a positive correlation with $T_{c,max}$ [9]. Computationally, the axial orbital energy and the range parameter of the intralayer hopping [12], the Madelung potential of the apical oxygen [13], and the charge-transfer energy between the in-plane Cu and oxygen [14], all have been shown to correlate with the material dependence of $T_{c,max}$. Recently, Kim *et al*. suggested the bond strength between apical atoms and the Bader charge of apical oxygen, which are related to phonons of the apical oxygen and out-of-$CuO_2$-plane charge fluxes, as material dependent parameters [15]. A generalized flux picture to further emphasize the importance of dynamic anharmonic coupling of phonons and fluxes was shown to be able to predict a few important experimental phenomena in the pseudogap and strange metal phases of cuprates [16].

Even though previous works have reported these important correlations between certain parameters and $T_c$, which helped the understanding of materials dependence, it often lacks a quantitative analysis and prediction of the material dependence of $T_c$ for the existing or non-existing structures. On the other hand, the data-driven machine learning (ML) technique has been developed to predict the properties of materials without demanding a pre-known exact mechanism [17–19]. Especially, to predict $T_c$, several studies have been reported using various ML algorithms and features for BCS-type [20–22], Fe-based [23–25], Cu-based [23], and a broad class of superconductors [23,26–32].

Features based on the chemical composition of materials are commonly used for ML modeling because of their accessibility. For example, Stanev *et al.* reported the random forest (RF) model predicting the superconductivity of materials including low-$T_c$, Cu-based, and Fe-based superconductors using chemical features [23]. However, the physical interpretability of the chemical composition features to the output prediction is not always directly understandable along with black-box algorithms of ML. Recently, Xie *et al.* suggested a ML model in a functional form with the electron-phonon coupling λ, logarithmic average phonon frequency $\omega_{\log}$, Coulomb pseudopotential $\mu^*$ for BCS-type superconductors [20]. Along this line, it would be worth exploring the direct relationship between material dependent parameters of cuprates and their $T_c$ using ML.

In this work, we aim to develop ML models for predicting $T_{c,max}$ of hole-doped cuprates using material dependent parameters as features, with their relation to $T_{c,max}$ having been reported. We select four features, including the Bader charge of apical oxygen, the bond strength between apical atoms, the number of superconducting layers, and the apical oxygen height, which are easily accessible from first-principles calculations and experimental crystal structures. To reveal a direct connection between the four features and the target variable, $T_{c,max}$, we search for analytical explicit equations by applying the linear regression with non-linear functional forms by the brute force searching (BFS) algorithm. The non-parametric RF is also employed to analyze the quantitative contribution of each feature to the output. Our model distinguishes from the previous ML models on the $T_c$ prediction by using the material dependent parameters beyond simple composition features and providing an explicit functional form with features for high $T_c$ cuprates. Furthermore, we generate hypothetical cuprates by changing the apical cation and predict the corresponding $T_{c,max}$ based on the developed BFS model. The cuprate with Ga as an apical cation is predicted to exhibit the $T_{c,max}$ that might be comparable to the Hg-family.

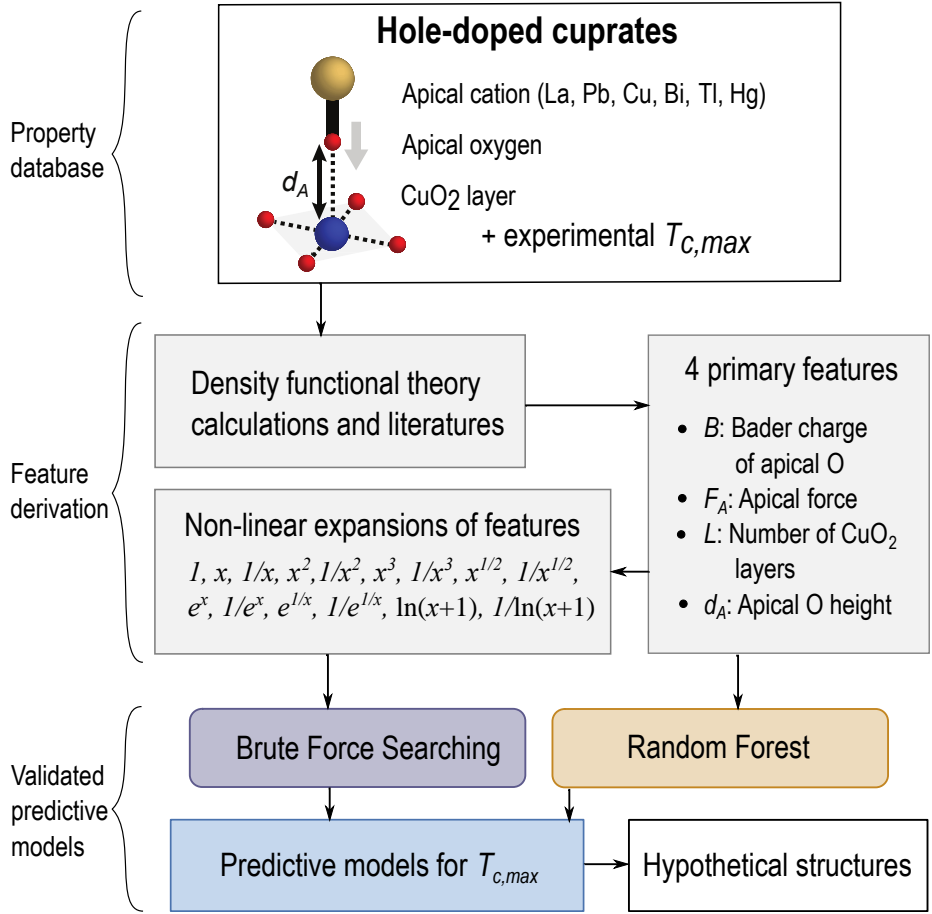

FIG. 1. Schematic workflow of the ML prediction models for $T_{c,max}$ of hole-doped cuprates

The workflow of this work is shown in Figure 1. All density functional theory calculations were performed with Vienna Ab inito Simulation Package (VASP) [33]. PAW-PBE functional was employed for the exchange-correlation functional [34]. We performed non-spin polarized calculations with $U = 8$ eV and $J = 1.34$ eV for the Cu $d$ orbitals [35] and with a 520 eV plane-wave energy cutoff. We used the four features of 29 hole-doped cuprates, which has been reported as material dependent parameters [9,15]; the Bader charge of the apical oxygen ($B$), the apical force ($F_A$), the number of CuO$_2$ layers ($L$), and apical oxygen height ($d_A$). Computational details and data acquisition are in the Supplemental Material [36].

We employed two machine learning algorithms; one is a parametric BFS model [37–40] and another is a nonparametric RF regression [41]. To avoid overfitting, we evaluated the performance of each model using leave-one-out cross-validation (LOOCV). The BFS model is a kind of linear regression with various functions converting primary features to the nonlinear functional forms of compound features. We have four primary features ($B$, $F_A$, $L$, $d_A$) and 15 functions; $1, x, \frac{1}{x}, x^2, \frac{1}{x^2}, x^3, \frac{1}{x^3}, \sqrt{x}, \frac{1}{\sqrt{x}}, e^x, \frac{1}{e^x}, e^{1/x}, \frac{1}{e^{1/x}}, \ln(x+1), \frac{1}{\ln(x+1)}$, which produce 57 distinct prototypical features. Multiplication of these prototypical features taken either two, three, or four generates 50,625 compound features. The compound features are used to construct linear regression models in the form of $T_{c,max} = \theta_0 + \sum_{i=1}^{m} \theta_i f_i$ where $\theta_i$, $f_i$, $m$ represent the regression coefficients, compound features, and the total number of compound features, respectively. We define $m$C$n$F model as a linear regression model with $m$ compound features where each compound feature consists of $n$ primary features. Note that each compound feature in a $m$C$n$F model can have different $n$ primary features. One example of a 3C2F model is $\theta_0 + \theta_1 \frac{e^B}{e^L} + \theta_2 \frac{1}{\sqrt{F_A L}} + \theta_3 \frac{e^{\frac{1}{L}}}{d_A}$. We searched the best BFS fit using all data as a training set and evaluated the three best-fit BFS models for each case using the LOOCV. On the other hand, the RF method is one of the most widely used ML algorithms due to its simplicity and ability to learn non-linear dependencies. Especially, the RF provides the importance of each feature, so-called Gini importance (GI). Details of machine learning algorithms are in the Supplemental Material [36].

Figure 2(a) and (b) show the performance of BFS models for predicting $T_{c,max}$ depending on the number of primary features and compound features in a model. In general, increasing the number of features improves the prediction performance. We consider up to three primary features (3F) in the model with two compound features (2C) because the model performances of the 4F is worse than those of the 3F in the 1C model. The most complicated model we developed is the 3C2F model, which has a similar but slightly better prediction performance than the 2C3F. The best model of the 3C2F exhibits the root-mean-square-error (RMSE) of 3.705K with the coefficient of determination, $R^2$ of 0.969 as in Fig. 1(c). As a comparison, a simple linear regression model with four primary features has the RMSE of 12.26 K and $R^2$ of 0.478. (See Supplemental Material for the results of the 2C3F model [36])

The best three BFS models for the 3C2F are summarized in Table 1 showing the empirical equation to predict the $T_{c,max}$ of cuprates. The best three models exhibit similar prediction performances according to the RMSE and $R^2$. The complicated equation with functionals such as $e^x$, $\sqrt{x}$, $\ln(x+1)$ indicates the non-linear relationship between features and $T_{c,max}$. Interestingly, the apical oxygen height ($d_A$) feature does not appear in the best three, which implies that the $d_A$ is less important than other features of Bader charge of the apical oxygen ($B$), apical force ($F_A$), and number of $CuO_2$ layers ($L$). It is worth noting that we can predict the $T_{c,max}$ and obtain an analytic functional form for $T_{c,max}$ of cuprates with high accuracy, which only needs the easily available three material dependent parameters.

To investigate the relationship between primary features and $T_{c,max}$, we have illustrated the shape of equation of the best 3C2F model with parameters of $B$, $F_A$, and $L$ as shown in Fig. 2(d). It shows that, for example, a material with $(B, F_A, L) = (7.3, 2.0, 2)$ expects to have a $T_{c,max}$ of 104 K. Despite the different equation forms, the three equations exhibit a similar shape as in Fig. S1, and their general shapes have common characteristics that (i) the predicted $T_{c,max}$ increases as $B$ decreases, (ii) predicted $T_{c,max}$ increases as $F_A$ increases, and (iii) the predicted $T_{c,max}$ has the maximum around $L$ of 3 or 4, which are consistent with the previous paper [15]. Our BFS model thus properly captures the expected relationship between each feature and $T_{c,max}$.

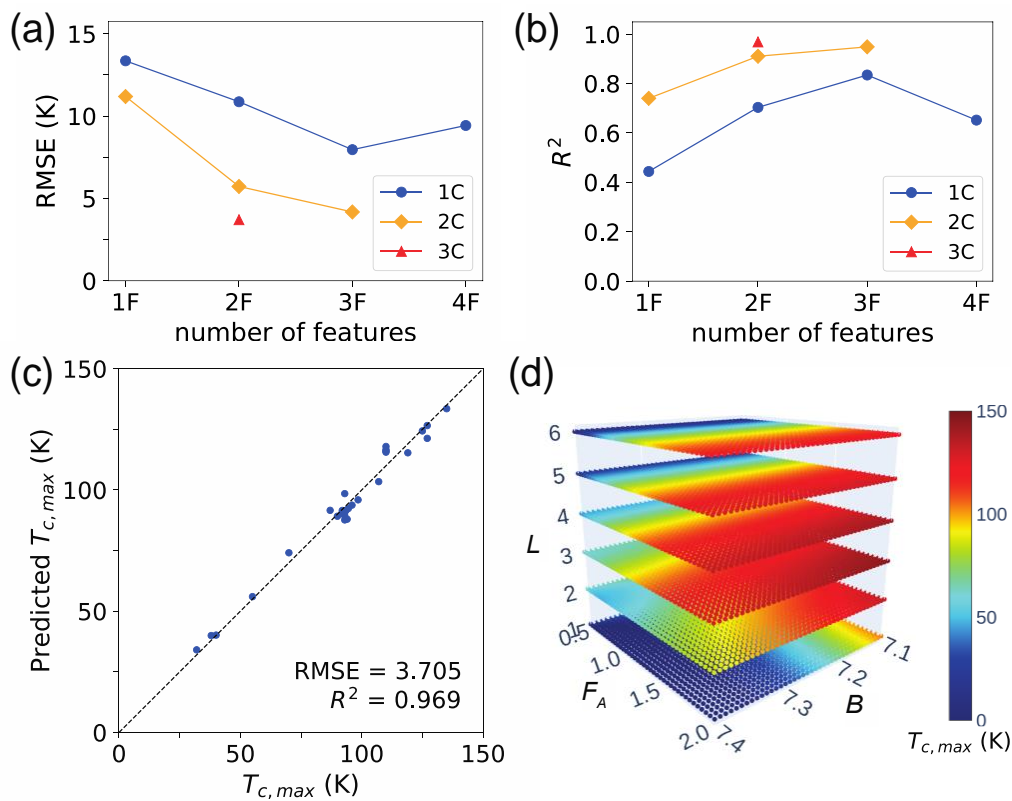

FIG. 2. Prediction performance of the BFS models. (a) Computed RMSE and (b) $R^2$ depending on the number of primary features and compound features. (c) Comparison of the experimental $T_{c,max}$ against the predicted $T_{c,max}$ obtained by the best 3C2F model. (d) The shape of the equation of the BFS model with the smallest RMSE of 3.705 K. The color bar indicates the calculated $T_{c,max}$.

Table 1 Details of model equations of the best three 3C2F models for prediction of $T_{c,max}$. $F_A$, $B$, and $L$ represent the apical force, the Bader charge of the apical oxygen, and the number of $CuO_2$ layers, respectively.

| | Model equations | RMSE (K) | $R^2$ |
|---|---|---|---|
| 1 | $96.3 - 0.789\dfrac{e^B}{e^L} + 689\dfrac{1}{\sqrt{F_A L}} - 348\dfrac{e^{\frac{1}{L}}}{e^{F_A}}$ | 3.705 | 0.969 |
| 2 | $-21.6 + 307\dfrac{10^3}{e^B\sqrt{L}} - 611\dfrac{\sqrt{F_A}}{e^L} + 58.1\dfrac{F_A^2}{L}$ | 3.795 | 0.957 |
| 3 | $-19.1 - 0.705\dfrac{e^B}{e^L} + 240\dfrac{\sqrt{F_A}}{\sqrt{L}} + 327\dfrac{1}{\ln(F_A+1)e^L}$ | 4.197 | 0.955 |

We have performed the nonparametric RF regression model, which can deal with complex non-linearity, to further investigate the importance of each feature to $T_{c,max}$. Figure 3(a) shows the performance of the RF depending on the number of features. Each point was obtained from the feature combination with the smallest RMSE. When only using two features; $B$ and $L$, the RF model exhibits the best performance with the RMSE of 7.735 K and $R^2$ of 0.837. The second-best RF model has the RMSE of 9.196 K and $R^2$ of 0.731 with three features of $B$, $L$, and $F_A$. The performance of the RF model is not better than the BFS model but shows a reasonable accuracy.

Table 2 represents the GI, which measures feature relevance on the output parameter, in our case, $T_{c,max}$. The GI for $B$ and $L$ in the best RF model are 0.555 and 0.445, respectively, which implies a similar contribution of $B$ and $L$ to predict $T_{c,max}$. In the second-best RF model, $B$ and $L$ exhibit a slightly larger GI than that of $F$ but all three have a similar GI of ~0.3. These results suggest that $B$ and $L$ are the most important features to predict $T_{c,max}$ followed by $F_A$. $d_A$ would be the least important feature among them. The high GI of the apical Bader charge $B$ here may be also related to the charge flux discussions regarding the transport property and pseudogap phase [15,16].

Note that the parametric model and the non-parametric model have their own advantages. The parametric BFS model provides high accuracy as well as analytical and explicit formulas for $T_{c,max}$, which cannot be obtained from the RF model because of its implicit regression process. The non-parametric RF algorithm has advantages of low computational cost and reasonable predictability, in our case, only with simple two features. In addition, the RF model provides the quantitative feature importance.

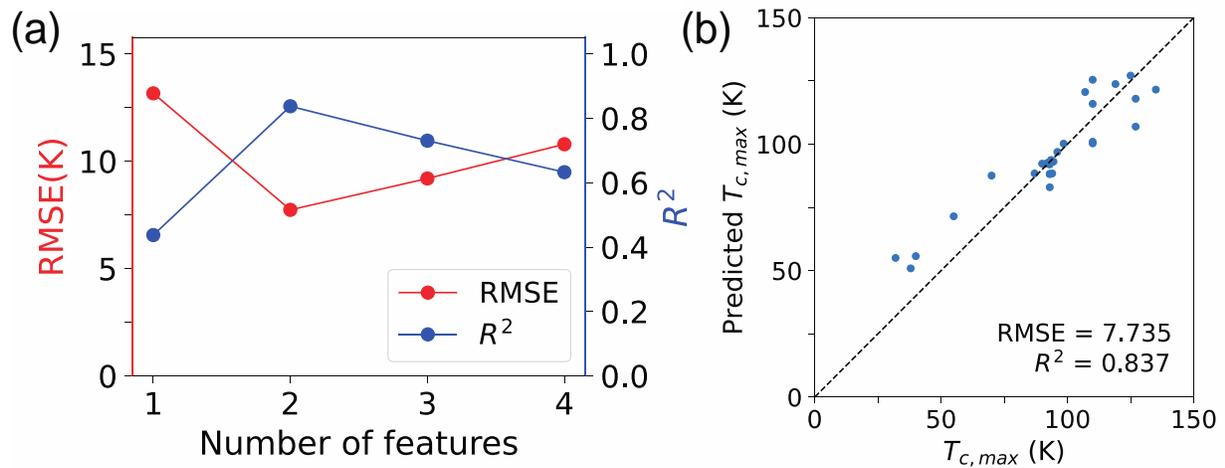

FIG. 3. Prediction performance of the RF models. (a) RMSE and $R^2$ according to the number of features. (b) Comparison of the experimental $T_{c,max}$ against the predicted $T_{c,max}$ obtained by the best RF model with $B$ and $L$ as features.

Table 2. The Gini importance of each feature

| feature rank | 2 features | 3 features |
|---|---|---|
| 1 | B  0.555 | B  0.342 |
| 2 | L  0.445 | L  0.335 |
| 3 |  | $F_A$ 0.323 |

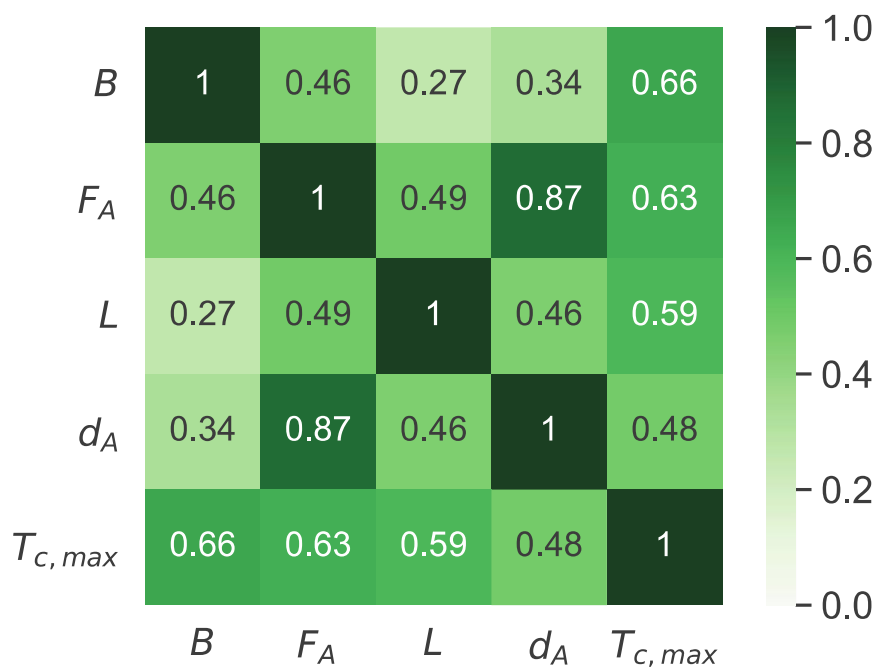

FIG. 4. The correlation map of the features and $T_{c,max}$ by the absolute values of the Pearson correlation coefficient.

We further investigate the correlation not only between a feature and $T_{c,max}$ but also among features using the Pearson correlation coefficient, $r_{xy}$. The Pearson correlation coefficient is defined as

$$r_{xy} = \frac{\sum_{i=1}^{m}(x_i - \bar{x})(y_i - \bar{y})}{\sqrt{\sum_{i=1}^{m}(x_i - \bar{x})^2}\sqrt{\sum_{i=1}^{m}(y_i - \bar{y})^2}}$$

, where $x_i$, and $y_i$ are parameters indexed with $i$ among the total data size, $m$. The upper bar indicates the average. The Pearson correlation coefficient shows the strength of linear correlation between two parameters in a range of -1 to 1. When the coefficient is close to |1|, it means a strong correlation.

Figure 4 shows the Pearson correlation map for the features and $T_{c,max}$. $B$, $F_A$, and $L$ have a strong and similar correlation with $T_{c,max}$ with the coefficient $\geq$ 0.59, which is consistent with our BFS, RF model, and GI importance. $B$ and $L$ have the least correlation, which explains reasonable accuracy in the RF model only using $L$ and $B$. We also can see $F_A$ is highly correlated with $d_A$ exhibiting the coefficient of 0.87, which explains the high predictability even without the $d_A$ feature. In addition, the coefficient between $F_A$ and $T_{c,max}$ (0.63) is larger than that of $d_A$ and $T_{c,max}$ (0.48), which suggests that, if we have to choose one feature out of $F$ and $d_A$, $F$ would be a better choice.

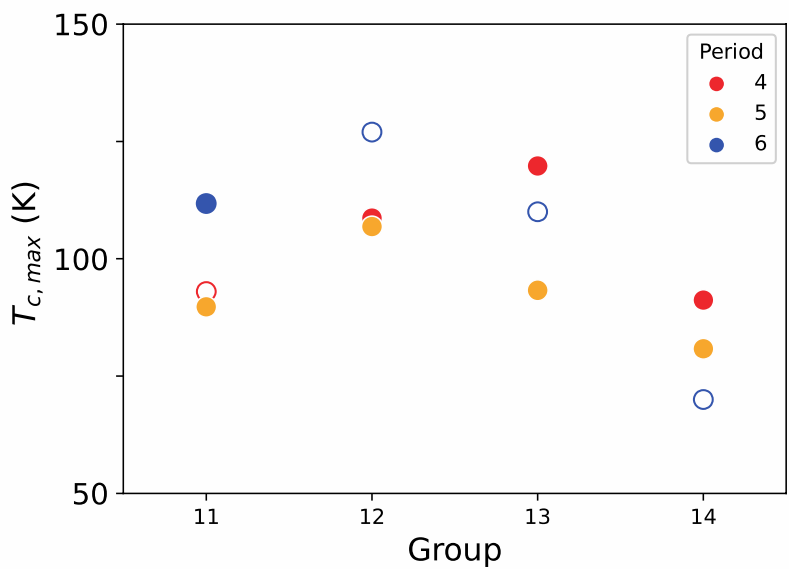

FIG. 5. Calculated $T_{c,max}$ of hypothetical structures using the BFS 3C2F models (solid circles). The group 11 (Cu) of period 4, Group 12 (Hg), 13 (Tl), and 14 (Pb) of period 6 are from the experimental $T_{c,max}$ (empty circles) [13,42–44]. The averaged values of the best three 3C2F models are used.

To search for a new cuprate candidate, we have generated hypothetical structures by replacing apical cations with other elements. Based on the HgBa$_2$CaCu$_2$O$_6$ (Hg-1212) structure, we changed the Hg atom with alkali-earth elements (Mg, Ca, Sr, Ba), elements in group 11 to 15 and in period 4 to 6. Figure 5 shows the average of predicted $T_{c,max}$ of the best three 3C2F models for hypothetical structures, compared with the experimental $T_{c,max}$ for YBa$_2$Cu$_3$O$_7$ (group 11, period 4), Hg-1212 (group12, period 6), Tl$_2$Ba$_2$CaCu$_2$O$_8$ (group13, period 6), and Pb$_2$Sr$_2$YCu$_3$O$_8$ (group14, period 6). For periods 5 and 6, $T_{c,max}$ is maximum with the group 12 elements as the apical cation. For period 4, $T_{c,max}$ is the highest with the group 13 element, Ga. Figure S3 shows the predicted $T_{c,max}$ using the BFS 2C3F model. Even though the estimated values are slightly different from those of the 3C2F model, the similar trend shows the robustness of the prediction [36].

We have investigated the Ga case, GaBa$_2$CaCu$_2$O$_6$ (Ga-1212), which has the highest $T_{c,max}$ among hypothetical structures and a comparable predicted $T_{c,max}$ to that of Hg-1212. The apical force and the Bader charge of Ga-1212 are 1.73 eV/Å and 7.20 while those of Hg-1212 are 1.66 eV/Å and 7.13, respectively. With the help of ML, the $T_{c,max}$ of new structures are quantitatively estimated, otherwise we can only speculate that Ga as the apical cation might have a chance to exhibit high $T_{c,max}$. We further generated and predicted the $T_{c,max}$ of GaBa$_2$Ca$_{n-1}$Cu$_n$O$_{2n+2}$ (Ga-12($n$-1)$n$, $n$ = 1, 2, and 3) and Ga$_2$Ba$_2$Ca$_{n-1}$Cu$_n$O$_{2n+4}$ (Ga-22($n$-1)$n$) as shown in Table 3. The base structure for the latter case is Tl$_2$Ba$_2$Ca$_{n-1}$Cu$_n$O$_{2n+4}$ (Tl-22($n$-1)$n$). The calculated $T_{c,max}$ for Ga-22($n$-1)$n$ is higher than those of Ga-12($n$-1)$n$. The $B$ of Ga-22($n$-1)$n$ and Ga-12($n$-1)$n$ are similar while the $F_A$ of Ga-22($n$-1)$n$, ~2.17 eV/Å is larger than that of Ga-12($n$-1)$n$, ~1.73 eV/Å, which results in the larger $T_{c,max}$ of Ga-22($n$-1)$n$. A few experimental studies reported the $T_c$ of cuprates having Ga as an apical anion (GaSr$_2$Ca$_{n-1}$Cu$_n$O$_{2n+3}$ type) whose $T_c$ are reasonably high as 49-68 K, 70-73 K, and 107 K for $n$ = 2, 3, 4, respectively [45–48]. It would be worth revisiting the Ga family with $GaBa_2Ca_{n-1}Cu_nO_{2n+2}$ type or $Ga_2Ba_2Ca_{n-1}Cu_nO_{2n+4}$ type structures, as our predictions show a higher $T_{c,max}$ in these two types.

Table 3 Predicted $T_{c,max}$ of Ga-12($n$-1)$n$ and Ga-22($n$-1)$n$ compared with experimental $T_{c,max}$ of Hg-12($n$-1)$n$. $n$ indicates the number of $CuO_2$ layers.

| $n$ | Ga-12($n$-1)$n$ (K) | Ga-22($n$-1)$n$ (K) | Hg-12($n$-1)$n$ (K) |
|---|---|---|---|
| 1 | 71 | 107 | 94 [49] |
| 2 | 117 | 142 | 127 [42] |
| 3 | 131 | 152 | 135 [50] |

In conclusion, we have developed machine learning models using the BFS and RF algorithms for predicting $T_{c,max}$ of hole-doped cuprates. To improve the physical interpretability of the models, we used four readily accessible materials dependent parameters from the DFT calculations ($F_A$ and $B$) and experimental structures ($L$ and $d_A$). The BFS model provides the empirical and explicit equation with three $B$, $F_A$, and $L$ parameters for $T_{c,max}$, which exhibits the high accuracy of the RMSE of 3.705 K and $R^2$ of 0.969. The RF model shows the reasonable accuracy of the RMSE of 7.735 K and $R^2$ of 0.837 only using two features of $B$ and $L$. We found that $B$, $L$, and $F_A$ features are important to predict $T_{c,max}$ by analyzing the BFS results, the GI importance, and Pearson's correlation coefficient. Furthermore, we predicted the $T_{c,max}$ of hypothetical structures where the Hg atom is replaced with other elements. When Ga is the apical cation, the $T_{c,max}$ are the highest with 71, 117, and 131 K for one, two, and three $CuO_2$ layers, respectively. We hope that this work can inspire the theoretical development for the $T_c$ equation and guide the experimental search of new cuprate superconductors.

**Acknowledgement**

We thank the helpful discussion with Hyungchul Kim and Xi Chen. This work was supported by the National Research Foundation of Korea (NRF) (Grant number 2019R1F1A1052026) and Korea Electric Power Corporation (Grant number R20XO02-12).